# Decoherence demystified : The hydrodynamic viewpoint

Kyungsun Na[*] and Robert E. Wyatt[†]
*Institute for Theoretical Chemistry, Department of Chemistry and Biochemistry*
*The University of Texas, Austin, TX 78712*

The hydrodynamic formulation of quantum mechanics is used to elucidate the mechanism for decoherence, the suppression of interference effects in a system evolving from an initial coherent superposition. Analysis of time-dependent trajectory ensembles, flux maps, and elements of the stress tensor for two composite systems, in one of which the system is uncoupled to the environment, leads to the decoherence mechanism. For the uncoupled case, the quantum force acting on the fluid elements directs flux toward an attractor where the interference feature arises. For the coupled case, the classical force acting on each fluid element counters the quantum force and leads to gradual separation of the components of the initial superposition. Concomitantly, fluid stress is relieved when flux vectors diverge from a repellor in the mid-region between the separating wavepackets, thus suppressing formation of the interference feature.

Decoherence is the physical process by which a superposition of quantum states evolves into a classical mixture where interference effects are prohibited [1-3]. The relationship between quantum and classical descriptions of processes, quantum measurement theory and quantum computation are some areas where decoherence continues to play a fundamental role. The hydrodynamic formulation of quantum mechanics provides a unique window for viewing the mechanism of this seemingly controversial process. Here we investigate the decoherence of a superposition of quantum states by comparing the dynamics of two composite systems, in one of which the system-environment coupling is turned off. The decoherence mechanism emerges through analysis of the time-dependent hydrodynamic fields, especially quantum trajectories for the fluid elements and the inter-related flux and stress maps for the quantum fluid.

It is generally accepted that decoherence occurs either when a quantum system interacts with a many degree of freedom heat bath or when it is entangled with an environment with a few degrees of freedom [2]. Conventionally, influence functionals [4], quantum master equations [5], and Wigner functions [6] have been used to predict and analyze the destruction of interference effects attributed to decoherence. The double-slit diffraction experiment has also fermented controversy relating to the disappearance of interference effects when particle paths are measured [7-9]. Ever since the Einstein-Bohr debate [10], this experiment has been said to evoke the basic *mystery* of quantum mechanics [11]. Also adding to the seemingly mysterious nature of decoherence is the statement [12]: 'It is difficult to give a simple explanation for the existence of decoherence…it occurs because one is dealing with an environment that is very complex, with *too many* degrees of freedom. Decoherence is therefore…very *difficult to obtain from a theoretical viewpoint*' ( italics added for emphasis).

In this study, we emphasize new insights gleaned from the hydrodynamic analysis of an interference experiment bearing analogies to the double-slit experiment. Initially, a coherent superposition of two well separated wave packets is prepared in a composite system involving a system mode, x, coupled to one harmonic bath mode, y. The Hamiltonian for the composite system is decomposed into system, (harmonic) bath and coupling contributions

$$H = H_s + H_b + H_c = \frac{p_x^2}{2m_0} + \frac{1}{2}\left[\frac{p_y^2}{m} + ky^2\right] + cxy$$

The coupling potential is bilinear in the coordinates [13] and the decoherence rate depends upon the coefficient 'c'. The initial wavefunction is a superposition of 'left and right' localized Gaussian functions for the system times a ground state harmonic oscillator function for the bath (N is a normalization factor)

$$\Psi(x,y,t=0) = \frac{1}{\sqrt{2}}\left[Ne^{-\beta(x-a)^2} + Ne^{-\beta(x+a)^2}\right]\varphi_0(y).$$

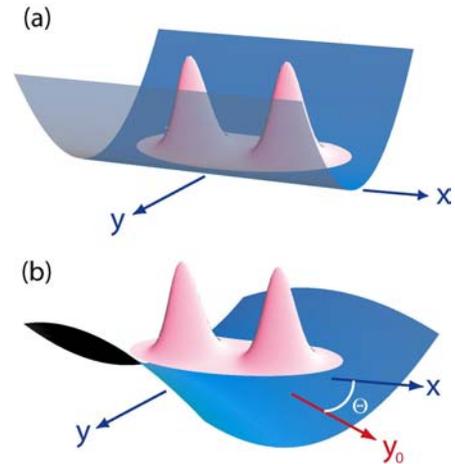

FIG. 1. Potential energy surface and the initial density distribution for two cases: (a) without and (b) with system-bath coupling. The axes x and y denote the system and bath coordinates, respectively. The direction $y_0$ runs along the valley floor for the coupled case. For the coupled case in (b), the x and $y_0$ axes make an angle $\vartheta = -19^0$. The initial wave packet is a coherent superposition for the system and a ground state harmonic oscillator function for the bath. The two superposed Gaussians shown here are initially centered on the x-axis.

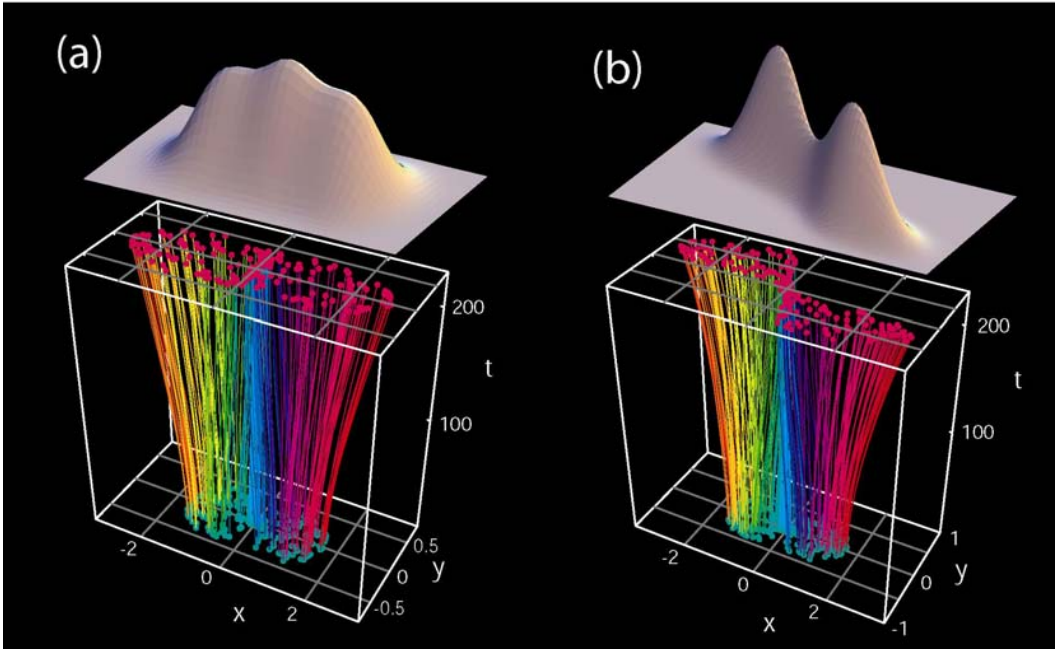

FIG. 2. Time evolution of 200 quantum trajectories from t=0 until t=225 and the probability density at t=225 for the (a) uncoupled and (b) coupled cases. Atomic units are used except that one time step is 2 a.u. Since the density along each of the quantum trajectories changes during the time evolution, the distribution of fluid elements does not represent the probability density directly. Density plots shown above the quantum trajectory plots illustrate coherence and decoherence in (a) and (b), respectively.

The parameters [14] in this wavefunction were chosen to insure small overlap between the two Gaussians, which are initially centered at $x = \pm a$.

Potential energy surfaces and initial wave packets are shown in Figure 1. For the uncoupled case (a), two separated Gaussians form the superposition while the potential allows for free motion in the x-direction and harmonic bath motion in the y-direction. For the coupled case (b), the potential exhibits two descending valleys as one moves away from the origin along the $y_0$-direction. (The potential for this case can be decomposed into a term for the uncoupled composite system plus the coupling term: $V=V_0+V_c$.)

The hydrodynamic formulation is initiated by substituting the polar form of the wavefunction, $\Psi(r,t) = R(r,t)\exp(iS(r,t)/\hbar)$, into the time-dependent wave equation [15-17]. We then obtain the continuity equation, $\dot{\rho} = -\rho \nabla \cdot v$, which connects the probability density, $\rho = R^2$, and the velocity by $v = j/\rho = (\nabla S)/m$, where $j$ is the probability flux. The second equation is $\dot{v} = -\nabla(V+Q)/m$, in which the flow acceleration is produced by the sum of the classical force, $f_c = -\nabla V$, and the quantum force, $f_q = -\nabla Q$. The dynamics is governed by the non-local quantum potential [17-18], $Q(r,t) = -(\hbar^2/2m)\nabla^2 R(r,t)/R(r,t)$ in addition to the classical potential, V. Finally, a dynamical equation for the action function is also obtained [19], $\dot{S} = 1/2m(\nabla S)^2 - (V+Q) = L_q$, where the quantum Lagrangian ($L_q$) measures the excess of the flow kinetic energy over the total potential. The hydrodynamic fields (position, velocity, probability density, and action function) along the trajectories are obtained by integrating these three dynamical equations. Unlike classical Hamiltonian flow, the quantum fluid is compressible (generally, $\nabla \cdot v \neq 0$), irrotational ($\nabla \times v = 0$) except near wavefunction nodes, and viscid (since the stress tensor, defined later, usually does not vanish).

In order to implement the hydrodynamic formulation, the initial wave packet is subdivided into N *fluid elements* [20]. Once we obtain the probability density and action functions computed along each trajectory, the wavefunction may be synthesized [19]. The wavefunction at space-time point (r, t) can be obtained from the wavefunction $\Psi(r_0,t_0)$ at $(r_0,t_0)$ along the trajectory linking these points:

$$\Psi(r,t) = \exp\{-\frac{1}{2}\int_{t_0}^{t}\nabla \cdot v d\tau - \int_{t_0}^{t}L_q(\tau)d\tau\}\Psi(r_0,t_0)$$

where the first exponential updates the amplitude and the argument of the second exponential updates the action function along the trajectory.

Evolution of the fluid elements was obtained [19,21-23] by integrating the equations of motion in the Lagrangian, moving with the fluid, picture. A significant advantage of this picture is that a small number of fluid elements may be required, especially for higher dimensional problems [24]. In this formulation, the initially structured grid develops into an irregular mesh as time goes on. Difficulties may arise when calculating derivatives on these sparse unstructured grids. For this purpose, we use the weighted moving least squares algorithm [21,25]. The function is expanded in a set of $n_b$ local basis functions $\{p_k(\xi,\eta)\}$, where $(\xi,\eta)$ denote displacements from the target point to the $n_p$ nearest neighbor points [26], $f(x,y) = \sum_{k=1}^{n_b} a_k(t) p_k(\xi,\eta)$. Once the expansion coefficients $\{a_k\}$ are found by solving a system of linear equations, the partial derivatives may be evaluated. In this study, we used the 10 term cubic basis set $\{1,\xi,\eta,\xi^2,\eta^2,\xi\eta,..\}$. During the time evolution, some of the fluid elements may be forced into close proximity. In order to counter this compression, the mesh was adapted on every time step. All fields needed at the next time step were then interpolated onto the new uniform mesh.

We first compare a set of quantum trajectories and the resultant probability density for the uncoupled and coupled cases in Figure 2 (at t=225 time steps). Streamlines followed by the fluid elements expand outward from the localized components of the initial distribution. For the uncoupled case (a), trajectories in the central region merge near the attractor, the mid-plane between the two separated initial wave packets. The probability density shown above the streamlines displays significant interference buildup due to the quantum force. When the system-bath coupling is turned on, see part (b), the classical force encourages the superposed states to split into components moving in opposite directions. However, the quantum force still plays a role by trying to push fluid elements toward the central region. Moderation of the quantum force by the classical force suppresses density buildup in the central region, thus leading to decoherence. The density plot shown above the trajectories illustrates suppression of the interference feature.

We now investigate a pair of inter-related hydrodynamic fields, the flux vector distribution and the stress tensor for the quantum fluid. In classical hydrodynamics, the Navier-Stokes (NS) equation [27] governing the change in the momentum density, $\rho \cdot mv$, is given by $\partial(\rho m v_i)/\partial t = -\sum_j \nabla_j \Pi_{j,i} - \rho \nabla_i V$, where the last term is the 'external' force density arising from the potential V and where $\Pi$ is the stress tensor (units of pressure, force/area, or momentum flux, momentum/(area·time)). The classical stress tensor contains both compressive (normal force) and shear (tangential force) terms. The shear tensor measures the strain rate resulting from 'squashing' a tiny box centered at a point in the fluid. From the quantum equations of motion for $\rho$ and $v$, a quantum version of the NS equation can be derived [28]. In this case, the stress tensor has both classical and quantum components (depending on $\hbar^2$), $\Pi_{i,j} = \Pi_{i,j}^c + \Pi_{i,j}^q$, with the classical part given by $\Pi_{i,j}^c = m\rho v_i v_j$. The quantum part has scalar pressure and compressive stress terms on the diagonal, and off-diagonal shear stress terms, $\Pi_{i,j}^q = P\delta_{i,j} + \Pi_{i,j}^{shear}$, where $P = (-\hbar^2/(4m))\nabla^2 \rho$. The shear terms are conveniently expressed in terms of the Einstein osmotic velocity, $u = -(D/\rho)\nabla\rho$, where the quantum diffusion coefficient is $D = \hbar/(2m)$. This diffusion coefficient and $u$ play a significant role in stochastic quantum mechanics [29]. In terms of the components of $u$, the shear part of $\Pi_{i,j}^q$ is given by $\Pi_{i,j}^{shear} = m\rho u_i u_j$. Combining the classical and quantum components, the stress tensor takes the compact form, $\Pi_{i,j} = P\delta_{i,j} + m\rho \operatorname{Re}(w_i w_j^*)$, where $w$ is the complex velocity, $w = v + iu$.

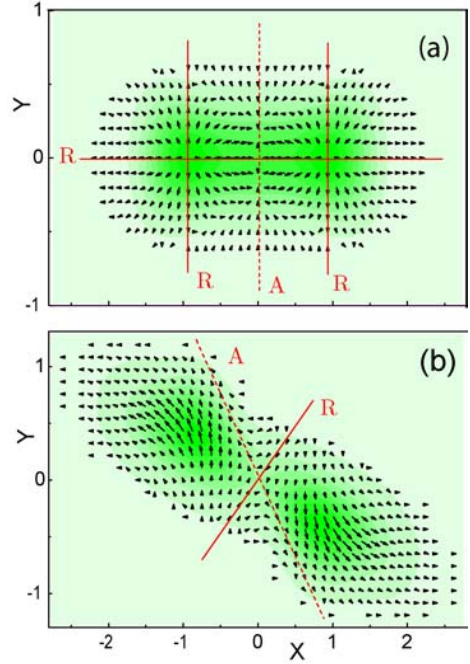

FIG. 3. Flux vector maps superimposed on contour maps of the stress tensor component $\Pi_{1,1}$ for the (a) uncoupled and (b) coupled cases. In (a), most of the flux vectors are pointed away from the high density and high stress regions toward regions of lower stress. The vertical midplane acts as an attractor, marked A, for flux vectors; this is where density builds up and the interference pattern forms. Flux vectors diverge from the three repellors, marked R, In (b), flux vectors are drawn toward the attractor (tilted dashed line) and diverge from the repellor (tilted solid line). The peaks of the two wavepackets move away from this repellor and gradually separate into the two valleys shown in Fig. 1 (b). This process suppresses formation of the interference feature which forms in the middle of part (a) of this figure, thus leading to decoherence.

At each time, $\Pi_{i,j}(x,y)$ contributes to changes in the momentum density. If a marble is placed at any point on a 3D surface portraying this function, it will roll downhill, searching out a minimum, so as to 'relieve the stress'. This mechanical analogy is doing just what the NS equation claims: momentum density vectors adjust according to the steepest downhill path on the stress surface. Given enough time, the stress surface tends to flatten out.

Flux vector distributions superimposed on the stress maps are illustrated in Fig. 3 at t=200. For the uncoupled case (a), most of the flux vectors point away from the high stress and high density regions near $x = \pm 1$. In the central region, this flux is directed toward the mid-plane, which acts as an attractor, and this results in development of the intense interference feature. Flux vectors diverge from the three repellors which are also indicated in this figure. Flux vectors for the coupled case (b) try to align themselves with the potential valleys. In the central region, flux vectors diverge from the repellor which is tilted from the lower left to the upper right. This flow away from the region near the origin leads to suppression of the interference feature noted in part (a).

For the system studied here, there are three independent elements of $\Pi$, denoted $\Pi_{0,0}$, $\Pi_{0,1}$, and $\Pi_{1,1}$. but we will only display the diagonal 'bath-bath' element $\Pi_{1,1}$. Figure 3 also shows contour maps of $\Pi_{1,1}$ for the uncoupled and coupled cases. This stress element contains three components: a classical term ($m\rho v_1^2$), the quantum pressure depending upon $\nabla^2 \rho$, and the quantum normal stress term ($m\rho u_1^2$). For the uncoupled case, the classical stress makes a negligible contribution to the total stress because $v_1^2 << u_1^2$. Consequently, the pressure term and the quantum stress totally dominate the features shown in (a). At this time step, the density is concentrated near the x-axis between x=-1 and x=+1, the osmotic velocity component is approximately linear in y and the quantum stress is therefore proportional to $y^2 \rho$. However, the pressure term is large near the x-axis so that the sum of the pressure and quantum stress terms also attains its largest value near this axis. For the coupled case shown in part (b), the pressure and classical stress terms dominate. The large value for the y-component of the velocity leads to the inequality $v_1^2 >> u_1^2$ and consequently the quantum stress plays a relatively small role. The stress is very large in the upper left and lower right of this figure; it is near these regions that the longest flux vectors are found. Again, we note that flux vectors are directed toward regions of lower stress, in accord with the quantum NS equation [30].

System-bath interplay registered by the reduced density matrix and Wigner function is described separately [22]. For the decoherent system, even when the system is coupled to only *one* bath oscillator, the off-diagonal element of the system density matrix is damped away and interference ripples lying between larger density peaks in the Wigner function tend to broaden and disappear as time goes on. Splitting of both position and momentum for the two separating components of the wave packet is also shown in the reduced density matrix and Wigner function for the bath.

In this study, comparisons of the time-dependent hydrodynamic fields for a system initiated in a coherent superposition, without and with coupling to a bath, demonstrated the decoherence mechanism. From the viewpoint of the quantum Navier-Stokes equation, the uncoupled and coupled systems differ through only one term, the coupling force density $-\rho \nabla V_c$ on the right side of this equation. This term plays a crucial role in the decoherence mechanism. For the uncoupled system, internal stress is relieved as the initially separated wave packets spread into each other, thus causing the interference feature near the attractor between the initially separated components of the superposition. However, for the coupled case, the force density arising from the system-bath coupling counteracts this tendency to relieve the quantum stress, thus suppressing formation of the interference feature. A unique role for this case is played by a repellor in the central region between the two initially separated wavepackets. Flux directed away from this repellor toward the valleys on the potential energy surface prevents buildup of the interference component of the density and thus leads to decoherence.

This research was supported in part by the Robert Welch Foundation and the National Science Foundation. We thank Justin Briggle for assistance with the graphics and the Texas Advanced Computing Center for providing access to the Cray SV1.

\* Electronic address: **na@physics.utexas.edu**
† Electronic address:**cman041@aurora.hpc.utexas.edu**